\documentclass{ws-procs9x6}

\begin{document}
\newcommand{\newc}{\newcommand}
\newc{\ra}{\rightarrow}
\newc{\lra}{\leftrightarrow}
\newc{\beq}{\begin{equation}}
\newc{\eeq}{\end{equation}}
\newc{\barr}{\begin{eqnarray}}
\newc{\earr}{\end{eqnarray}}

\title{DARK MATTER IN THE COSMOS-\\ EXPLOITING THE SIGNATURES OF ITS INTERACTION WITH NUCLEI}
\author{J.D. VERGADOS}
\address{Physics Department, University of Ioannina, Gr 451 10, Ioannina, Greece
$^*$E-mail:vergados@cc.uoi.gr}
%
%
\begin{abstract}
We review  various issues related to the direct detection of constituents of dark matter, which are assumed to be Weakly Interacting Massive Particles (WIMPs). We specifically consider heavy WIMPs such as: 1) The lightest supersymmetric particle LSP or neutralino. 2) The lightest Kaluza-Klein particles in theories of extra dimensions and 3) other extensions of the standard model. In order to get the event rates one needs information about the structure of the nucleon as well as as the structure of the nucleus and the WIMP velocity distribution. These are also examined
 Since the expected event rates for detecting the recoiling nucleus are extremely low and the signal does not have a characteristic signature to discriminate them against background we consider some additional aspects of the WIMP nucleus interaction, such as the periodic behavior of the rates due to the motion of Earth (modulation effect). Since, unfortunately, this is characterized by a small amplitude we consider other options such as directional experiments, which  measure not only the energy of the recoiling nuclei but their direction as well. In these, albeit hard, experiments one can exploit two very characteristic signatures: a)large asymmetries and b) interesting modulation patterns. Furthermore we  extended our study to include evaluation of the rates for other than recoil searches such as: i) Transitions to excited states, ii) Detection of recoiling electrons produced during the neutralino-nucleus interaction and iii) Observation of hard X-rays following the de-excitation of the ionized atom. 
\end{abstract}


\bodymatter
\section{Introduction}
\label{sec:1}
The combined MAXIMA-1 \cite{MAXIMA-1} ,  BOOMERANG \cite{BOOMERANG} , 
DASI \cite{DASI} ,  COBE/DMR Cosmic Microwave Background (CMB)
observations \cite{COBE} ,  the recent WMAP data \cite{SPERGEL} and
SDSS
 \cite{SDSS} imply that the
Universe is flat \cite{flat01} and that most of the matter in
the Universe is dark, i.e. exotic.These results have been confirmed and improved
by the recent WMAP data \cite{WMAP06}. The deduced cosmological expansion is
consistent with  the luminosity distance as a function of redshift of distant supernovae 
\cite{supernova1,supernova2,supernova3}.
According to the scenario favored by the observations 
there are various contributions to the energy content of our Universe.
The most accessible energy component is baryonic matter, which accounts for
$\sim 5\%$ of the total energy density. A component that has
not been directly observed is cold dark matter (CDM)): a pressureless fluid that is
responsible for the growth of cosmological perturbations through
gravitational instability. Its contribution to the total energy density is
estimated at 
$\sim 25\%$. The dark matter is expected to become more abundant 
in extensive halos, that
stretch up to 100--200 kpc from the center of galaxies.
The component with the biggest contribution to the energy density has
an equation of state similar to that of a cosmological constant and is characterized as dark energy. 
The ratio $w=p/\rho$ is negative and close to $-1$. 
This component is responsible for $\sim 70\%$ of the total energy density 
and induces
the observed acceleration of the Universe \cite{supernova1}$^-$\cite{supernova3} . 
The total energy density of our Universe 
is believed to take the critical value consistent with spatial flatness.
Additional indirect information about the existence of dark matter
comes from  the rotational curves \cite{Jung} . The rotational velocity of an object increases so long
is surrounded by matter. Once outside matter the velocity of rotation drops as the square root  of the 
distance. Such observations are not possible in our own galaxy. The observations of other galaxies, 
similar to our own, indicate that the rotational velocities of objects outside the luminous matter
do not drop. So there must be a halo of dark matter out there.

Since the non exotic component cannot exceed $40\%$ of the CDM
~\cite {Benne} , there is room for exotic WIMP's (Weakly
Interacting Massive Particles).\\
  In fact the DAMA experiment ~\cite {BERNA2} has claimed the observation of one signal in direct
detection of a WIMP, which with better statistics has subsequently
been interpreted as a modulation signal \cite{BERNA1} . These  data,
however, if they are due to the coherent process, are not
consistent with other recent experiments, see e.g. EDELWEISS and
CDMS \cite{EDELWEISS} . It could still be interpreted as due to the
spin cross section, but with a new interpretation of the extracted
nucleon cross section. 

Since the WIMP  is expected to be very massive, $m_{\chi} \geq 30 GeV$, and
extremely non relativistic with average kinetic energy $T \leq 100 KeV$,
it can be directly detected mainly via the recoiling
of a nucleus  in the WIMP-nucleus elastic scattering.

The above developments are in line with particle physics considerations.
\begin{enumerate} 
\item Dark matter in supersymmetric theories\\
The lightest supersymmetric particle (LSP) or neutralino
 is the most natural WIMP candidate. In the most favored scenarios the
LSP can be simply described as a Majorana fermion, a linear 
combination of the neutral components of the gauginos and Higgsinos
\cite{Jung}$^-$\cite{Hab-Ka} . 

 In order to compute the event rate one needs
 an effective Lagrangian at the elementary particle 
(quark) level obtained in the framework of supersymmetry 
~\cite{Jung,ref2,Hab-Ka} .
 One starts with   
representative input in the restricted SUSY parameter space as described in
the literature, e.g. Ellis {\it et al} \cite{EOSS04} , Bottino {\it et al} , 
Kane {\it et al} , Castano {\it et al} and Arnowitt {\it et al} \cite{ref2} as well as elsewhere
\cite{GOODWIT}$^-$\cite{UK01} .
We will not, however, elaborate on how one gets the needed parameters from
supersymmetry. 

Even though the SUSY WIMPs have been well studied, for tor the reader's convenience we will give a description  in sec. \ref{sec:diagrams} of the basic 
SUSY ingredients needed to calculate LSP-nucleus scattering
 cross section.

\item  Kaluza-Klein (K-K) WIMPs. \\
These arise in  extensions of the standard model with compact extra dimensions. In
such models  a tower of massive particles appear as Kaluza-Klein
excitations.
In this scheme the ordinary particles are associated with the zero
modes and are assigned K-K parity $+1$.
In models with Universal Extra Dimensions one can have
cosmologically stable particles in the excited modes because of a
discreet symmetry yielding  K-K parity $-1$ (see previous work \cite{ST02a,ST02b,CFM02} as well as
the recent review by Servant
\cite{SERVANT}).
 \\The kinematics involved is similar to
that of the neutralino, leading to cross sections which are
proportional $\mu^2_r$, $\mu_r$ being the WIMP-nucleus reduced
mass. Furthermore the
 nuclear physics input
  is independent of the WIMP mass, since for heavy WIMP $mu_r\simeq Am_p$.
   There are appear two differences compared to the
  neutralino, though,  both related to its larger mass.
  \\i) First the density
  (number of particles per unit volume) of a WIMP
 falls inversely proportional to its mass. Thus,
  if the WIMP's considered are much heavier than the nuclear
 targets, the corresponding event rate takes the form:
  \beq
 R(m_{WIMP})=R(A)\frac{A \mbox{ GeV
}}{m_{WIMP}}
 \label{eq:rate}
 \eeq
 where $R(A)$ are the rates extracted from experiment up to WIMP
 masses of the order of the mass of the target.
\\ii) Second the average WIMP energy is now  higher. In fact one
finds that $\langle T_{WIMP}\rangle =\frac{3}{4}m_{WIMP}
\upsilon^2_0\simeq 40 \left ({m_{WIMP}}/({100 \mbox{
GeV)}}\right )$keV ($\upsilon_0\simeq 2.2\times 10^5$km/s).
  Thus for a K-K WIMP with mass $1$ TeV, the average
WIMP energy is $0.4$ MeV. Hence, due to the high velocity tail of
the velocity distribution,  one expects {\bf an energy transfer to
the nucleus  in the MeV region. Thus many nuclear
 targets can now be excited by the WIMP-nucleus interaction and the de-excitation photons
 can be detected.}

 \end{enumerate}
In addition to the particle model one needs the following  ingredients:
\begin{itemize}
\item A procedure in going from the quark to the nucleon level, i.e. a quark 
model for the nucleon. The results depend crucially on the content of the
nucleon in quarks other than u and d. This is particularly true for the scalar
couplings as well as the isoscalar axial coupling ~\cite{Dree}$^-$\cite{Chen} . Such topics will be discussed in sec.
\ref{sec:nuc}.
\item computation of  the relevant nuclear matrix elements~\cite{Ress}$^-$\cite{SUHONEN03}
using as reliable as possible many body nuclear wave functions.
By putting as accurate nuclear physics input as possible, 
one will be able to constrain the SUSY parameters as much as possible.
The situation is a bit simpler in the case of the scalar coupling, in which
case one only needs the nuclear form factor.
\item  Convolution with the LSP velocity Distribution.
To this end we will consider here  Maxwell-Boltzmann \cite {Jung} (MB) velocity distributions, with an upper velocity cut off put in by hand. The characteristic velocity of the M-B distribution can be increased by
a factor $n$ ($\upsilon_0\rightarrow n \upsilon_0,~n\ge1 $)by considering the interaction of dark matter and dark energy \cite{TETRVER06}.

Other distributions are possible, such as
  non symmetric ones,  like those 
of Drukier \cite {Druk} and Green \cite{GREEN02} , or non isothermal ones, e.g. those arising from late in-fall of   
dark matter into our galaxy, like  Sikivie's caustic rings \cite {SIKIVIE} . In any event in a proper treatment the velocity distribution ought to be consistent with the dark matter density as, e.g., in the context of the Eddington theory \cite{OWVER} .
\end{itemize}


 Since the expected rates are extremely low or even undetectable
with present techniques, one would like to exploit the characteristic
signatures provided by the reaction. Such are: 
\begin{enumerate}
\item The modulation
 effect, i.e the dependence of the event rate on the velocity of
the Earth 
\item The directional event rate, which depends on the
 velocity of the sun around the galaxy as well as the the velocity
of the Earth. 
has recently begun to appear feasible by the planned experiments
\cite {UKDMC,DRIFT} .
\item Detection of signals other than nuclear recoils, such as
\begin{itemize}
\item Detection of $\gamma$ rays following nuclear de-excitation, whenever possible \cite{eji93,VQS04} .
This seems to become feasible for heavy WIMPs especially in connection with modified M-B distributions due to 
the coupling of dark matter and dark energy ($\langle T_{WIMP} \rangle \simeq n^2  40 \left ({m_{WIMP}}/({100 \mbox{
GeV})}\right ),~n\ge 1$keV)
\item Detection of ionization electrons produced directly in the LSP-nucleus collisions \cite{VE05,MVE05} . 
\item Observations of hard X-rays produced\cite{EMV05} , when the inner shell electron holes produced as above are filled.
\end{itemize}
\end{enumerate}

 In all calculations we will, of course, include an appropriate 
nuclear form factor and take into account
the influence on the rates of the detector energy cut off.
We will present our results a function of the
LSP mass, $m_{\chi}$, in a way
which can be easily understood by the experimentalists.


\section{The Feynman Diagrams Entering the Direct Detection of WIMPS.}
\label{sec:diagrams}
\subsection{The Feynman Diagrams involving the neutralino}
\label{sec:FEYLSP}
 The neutralino is perhaps the most viable WIMP candidate and has been extensively 
studied (see, e.g., our recent review \cite{JDV06}). Here we will give a very brief 
summary of the most important aspects entering the direct neutralino searches.
 In currently favorable supergravity models the LSP is a linear
combination~\cite{Jung} of the neutral four fermions 
${\tilde B}, {\tilde W}_3, {\tilde H}_1$ and ${\tilde H}_2$ 
which are the supersymmetric partners of the gauge bosons $B_\mu$ and
$W^3_\mu$ and the Higgs scalars
$H_1$ and $H_2$. Admixtures of s-neutrinos are expected to be negligible.
The relevant Feynman diagrams involve Z-exchange, s-quark exchange and Higgs 
exchange.
\subsubsection{The Z-exchange contribution.}
\label{sec:Z-exc}
\begin{figure}
\psfig{file=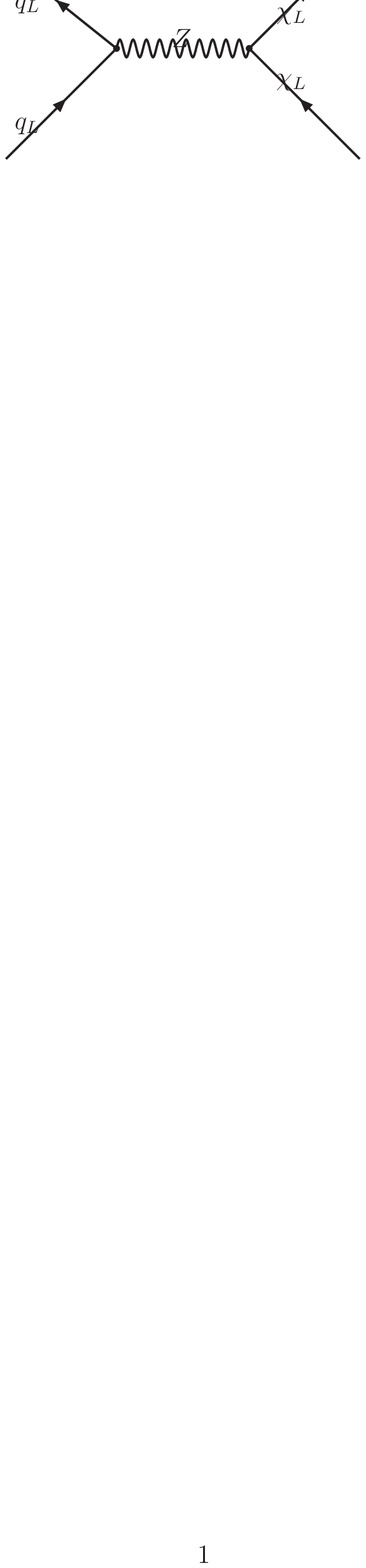,width=2.0in}
\psfig{file=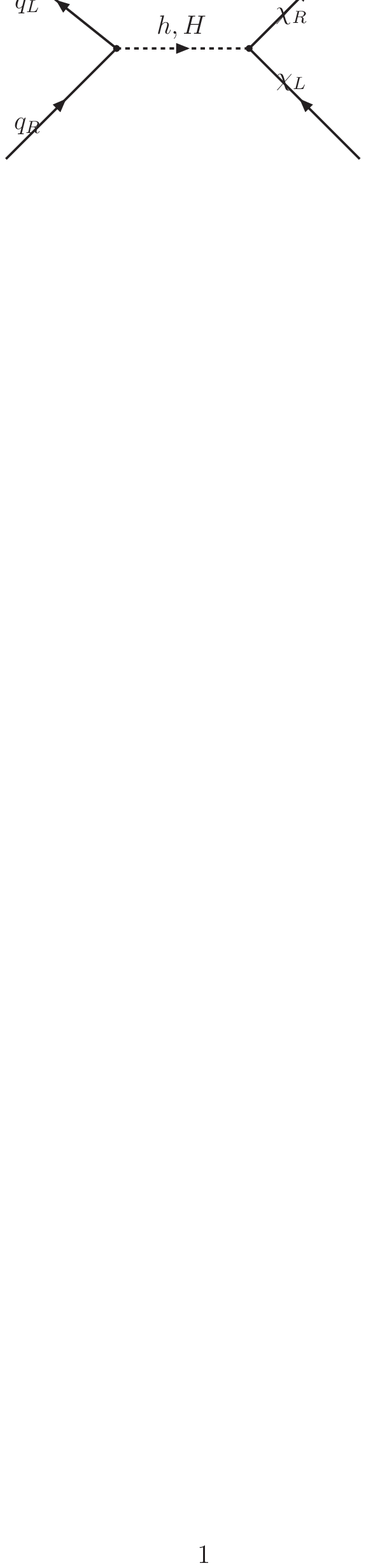,width=2.0in}
\caption{The LSP-quark interaction mediated by Z and Higgs exchange.}
\label{LSPZH}
\end{figure}

The relevant Feynman diagram is shown in Fig. \ref{LSPZH}. It does not lead to coherence, since $\bar{\Psi}\gamma_{\lambda}\Psi=0$ for a Majorana fermion like the neutralino (the Majorana fermions do not
have electromagnetic properties). The coupling $\bar{\Psi}\gamma_{\lambda}\gamma_5\Psi$ yields negligible contribution for a non relativistic particle in the case of the spin independent cross section \cite{JDV96}. It may be important in the case of the spin contribution, which arises
from the axial current).
\subsubsection{The $s$-quark Mediated Interaction }
\label{sec:sq-exc}
\begin{figure}
\begin{center}
\psfig{file=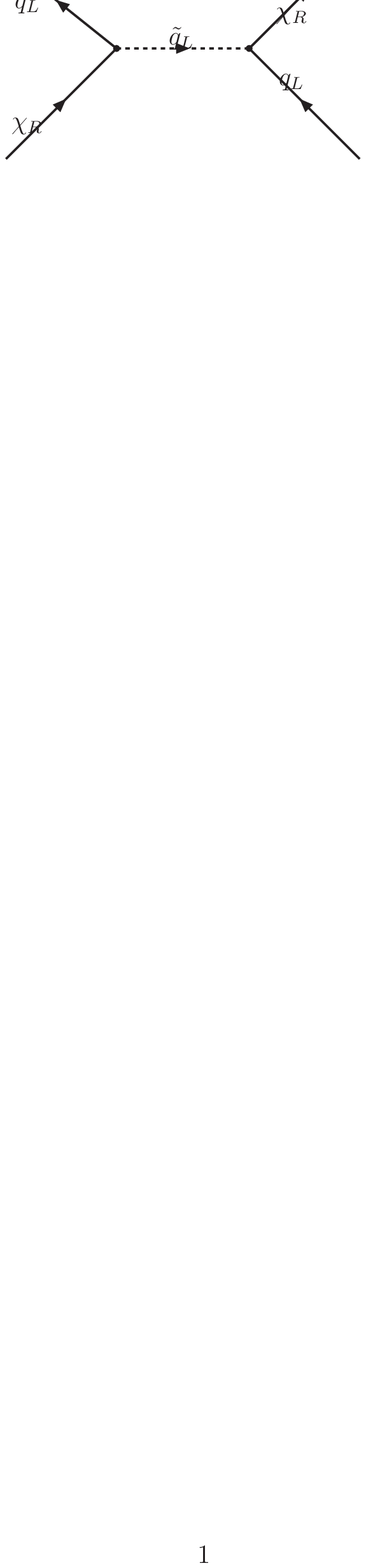,width=2.0in}
\psfig{file=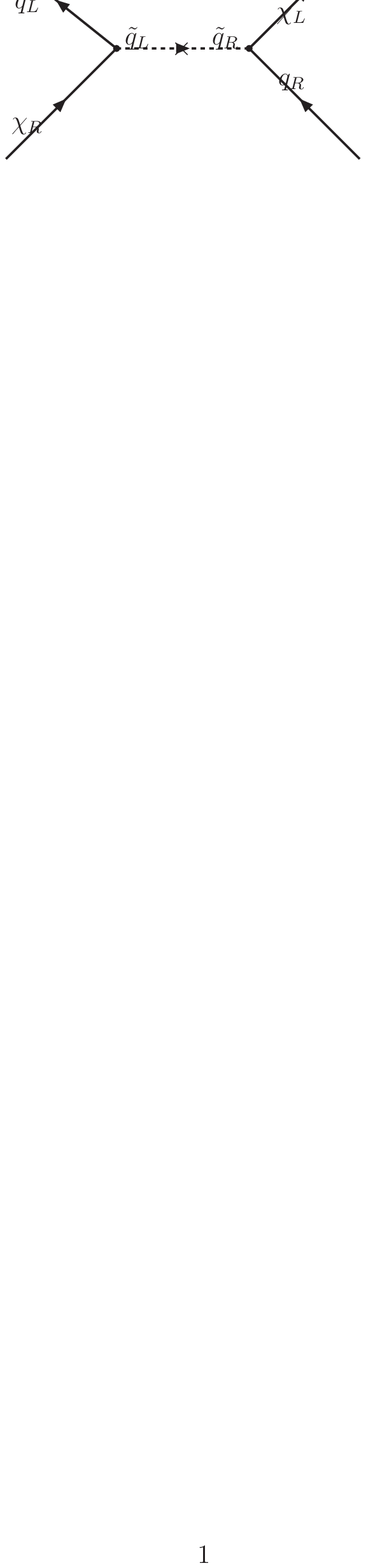,width=2.0in}
\end{center}
\caption{The LSP-quark interaction mediated by s-quark exchange. Normally it yields V-A interaction 
which does not lead to coherence at the nuclear level. If, however, the isodoublet
 s-quark is admixed with isosinglet one to yield a scalar interaction at the quark level.}
\label{LSPSQVS}
\end{figure}

The other interesting possibility arises from the other two components of
$\chi_1$, namely ${\tilde B}$ and ${\tilde W}_3$. Their corresponding
couplings to $s$-quarks (see Fig. \ref{LSPSQVS} ) can be read from the appendix C4 of Ref.~\cite{Hab-Ka} and our earlier review \cite{JDV06}.
Normally this contribution yields vector like contribution, i.e it does not lead to coherence. If,
however, there exists mixing between the s-quarks with  isospin $1/2$ ($\tilde{q}_L$) and the isospin 0 ($\tilde{q}_R$), the s-quark exchange 
may lead to a scalar interaction at the quark level and hence to coherence over all nucleons
at the nuclear level \cite{JDV06}.
\subsubsection{The Intermediate Higgs Contribution}
\label{sec:Higgs-exc}
The most important contribution to coherent scattering can be achieved via the intermediate Higgs
particles which survive as physical particles. In supersymmetry there exist two such physical
Higgs particles, one light $h$ with a mass $m_h\leq$120 GeV and one heavy $H$ with mass $m_H$, which
 is much larger. The relevant interaction can arise out of the
Higgs-Higgsino-gaugino interaction \cite{JDV06} leading to a Feynman diagram shown in Fig.
\ref{LSPZH}. 

In the case of the scalar interaction the resulting amplitude is proportional to the quark mass.
 \subsection{The Feynman Diagrams involving the K-K WIMPs}
\label{sec:FEYKK}
 \subsubsection{The Kaluza-Klein Boson as a dark matter candidate}
 \label{KK}
 We will assume that the lightest exotic particle, which can serve as a dark matter candidate, is a gauge boson $B^{1}$
 having the same quantum numbers and couplings with the standard model gauge boson $B$, except that it has K-K parity
 $-1$. Thus its couplings must involve another negative K-K parity particle. In this work we will assume that such  a particle can be one of
  the K-K quarks, partners of the ordinary quarks, but much heavier  \cite{ST02a,ST02b,CFM02} .
 \begin{itemize}
\item Intermediate K-K quarks.\\
this case the relevant
Feynman diagrams are
 shown in fig. \ref{fig:kkq}.
 \begin{figure}
\psfig{file=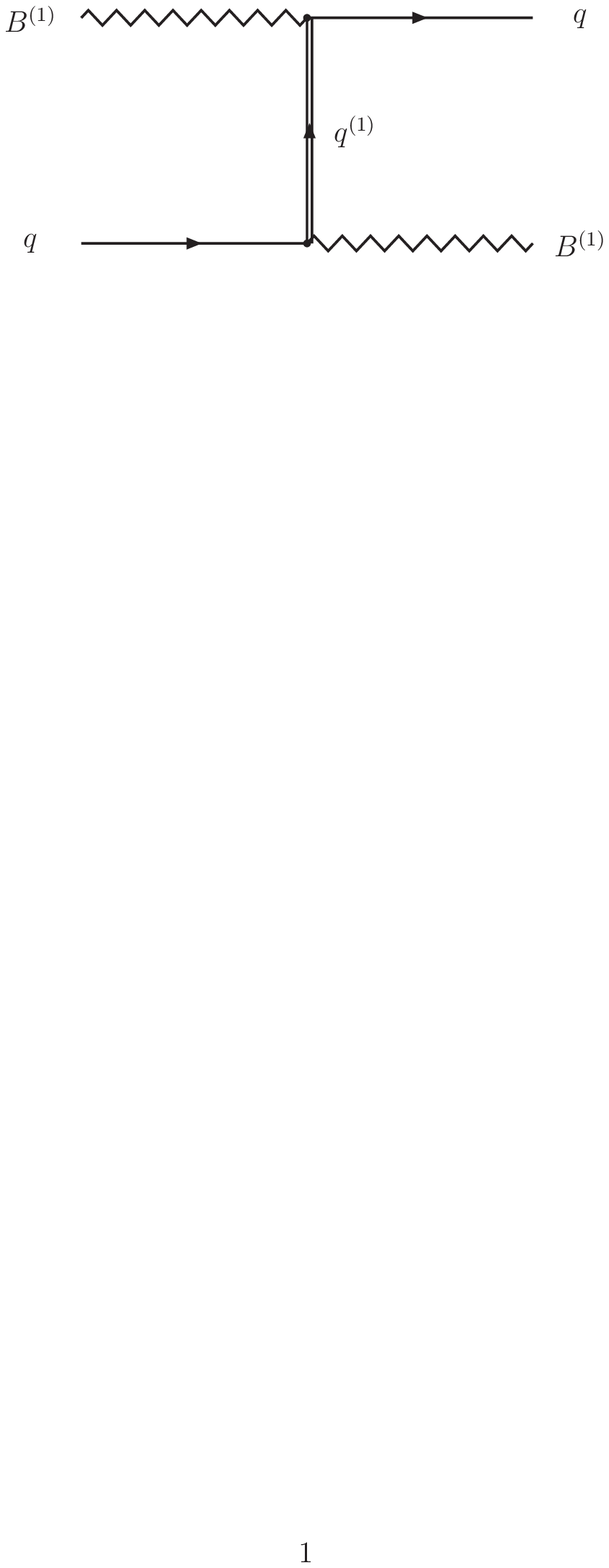,width=2.2in}
\psfig{file=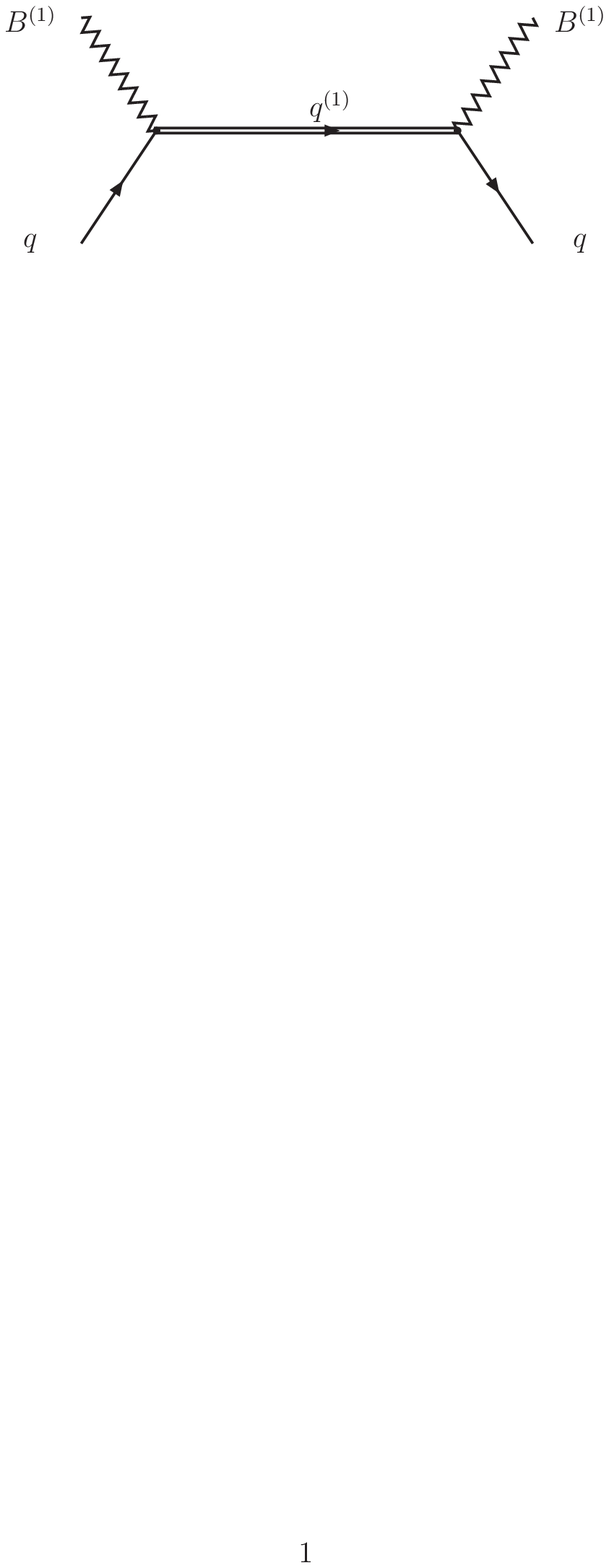,width=2.2in}
 \caption{K-K quarks mediating the interaction of K-K gauge boson $B^{1}$ with quarks at tree
level.}
 \label{fig:kkq}
  \end{figure}
\\The amplitude at the nucleon level can be written as:
 \beq {\cal M}_{coh}= \Lambda(\mbox{\boldmath
$\epsilon^{*'}$}.\mbox{\boldmath $\epsilon$})N\left [
\frac{11+12\tau_3}{54} \frac{m_p
m_W}{(m_{B^{(1)}})^2} f_1(\Delta )+\frac{1+\tau_3}{3} \frac{m_W}{
m_{B^{(1)}}} f_2(\Delta ) \right ] N \eeq
$$\Lambda=i 4 \sqrt{2} G_F m_W \tan^2{\theta_W
},f_1(\Delta )=\frac{1+\Delta +\Delta ^2 /2}{\Delta ^2(1+\Delta
/2)^2},$$
$$ f_2 (\Delta )=\frac{1+\Delta }{\Delta (1+\Delta /2)}~,
~\Delta =\frac{m_{q^{(1)}}}{m_{B^{(1)}}}-1$$ We see that the
amplitude is very sensitive to the parameter $\Delta $ ("resonance
effect").

In going from the quark to the nucleon level the best procedure is
to replace the quark energy by the constituent quark mass $\simeq
1/3m_p$, as opposed to adopting  \cite{ST02a,ST02b,CFM02} a procedure related to
the current mass encountered in the neutralino case \cite{JDV06}.


In the case of the spin contribution we find at the nucleon level
that:
\barr {\cal M}_{spin}&=& -i 4 \sqrt{2} G_F m_W \tan^2{\theta_W
}\frac{1}{3} \frac{m_p m_W}{(m_{B^{(1)}})^2} f_1(\Delta )
i(\mbox{\boldmath $\epsilon^{*'}$}\times \mbox{\boldmath
$\epsilon$}).
\nonumber\\
&&\left [ N\mbox{\boldmath $\sigma$} (g_0+g_1 \tau_3) N
\right ] \earr
$$g_0=\frac{17}{18}\Delta u+\frac{5}{18} \Delta d+\frac{5}{18} \Delta s~,
~g_1=\frac{17}{18}\Delta u-\frac{5}{18} \Delta d$$
for the isoscalar and isovector quantities \cite{JDV06}. The
quantities $\Delta_q$ are given by \cite{JDV06}
$$\Delta u=0.78\pm 0.02~,~\Delta d=-0.48\pm 0.02~,~\Delta s=-0.15\pm 0.02$$
We thus find $g_0=0.26~,~g_1=0.41\Rightarrow
a_p=0.67~,~a_n=-0.15$.
\item Intermediate Higgs Scalars.\\
The corresponding
Feynman diagram is shown in Fig. \ref{fig:kkhz}
   \begin{figure}[!ht]
\psfig{file=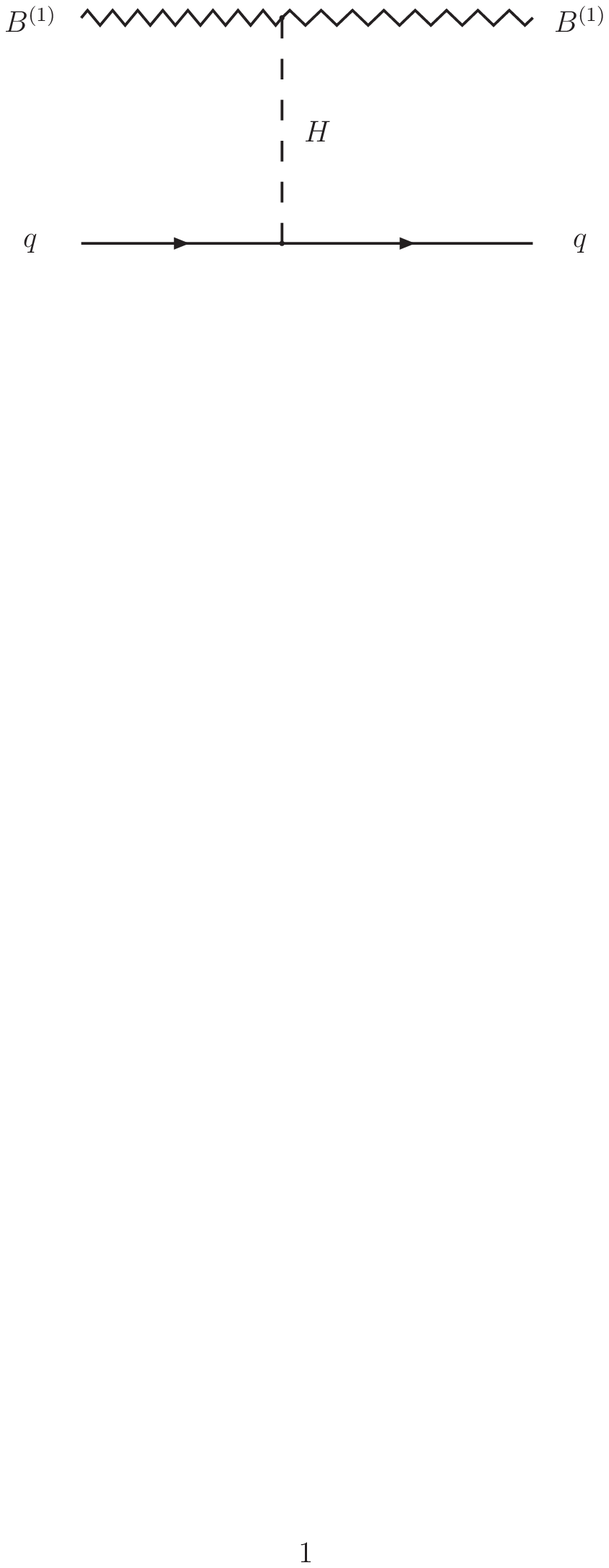,width=2.2in}
\psfig{file=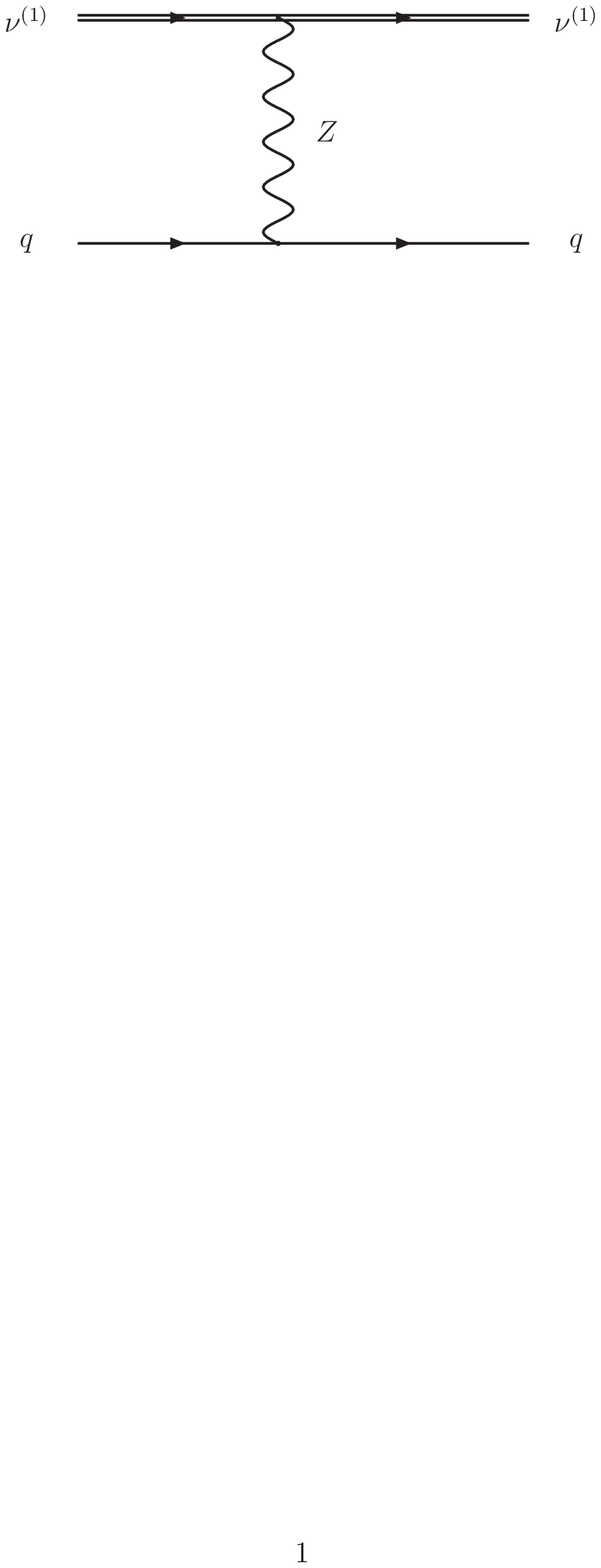,width=2.2in}
 \caption{The Higgs H mediating interaction of K-K gauge boson $B^{1}$ with quarks at tree
level (on the left). The Z-boson mediating the interaction of K-K
neutrino $\nu^{(1)}$ with quarks at tree level (on the right).}
 \label{fig:kkhz}
  \end{figure}
The relevant amplitude is given by:
  \beq
 {\cal M}_N(h)= -i~4 \sqrt{2}G_F m^2_W \tan^2{\theta_W}~\left [\frac{1}{4}\frac{m_p}{m^2_h}
 \left (-\mbox{\boldmath
$\epsilon^{*'}$}.\mbox{\boldmath $\epsilon$} \right ) \prec
N|N\succ  \sum_q f_q\right ]
  \eeq
   In going from the quark to the nucleon level we follow a procedure analogous to that
 of the of the neutralino,
  i.e.
  $\prec  N |m_q q \bar{q}|N  \succ  \Rightarrow f_q m_p$
\end{itemize}
  \subsubsection{K-K neutrinos as dark matter candidates}
  The other possibility is the dark matter candidate to be a heavy K-K neutrino.
   We will distinguish the following cases:
\begin{itemize}
\item Process mediated by Z-exchange.\\
 The amplitude associated with the diagram of Fig. \ref{fig:kkhz} becomes:
  \beq
{\cal M}_{\nu^{(1)}}=-\frac{1}{2 \sqrt{2}}G_FJ^{\lambda}(\nu^{(1)}) J_{\lambda}(NNZ)
 \eeq
 with $J_{\lambda}(NNZ)$ the standard nucleon neutral current and
$$J_{\lambda}(\nu^{(1)})= \bar{\nu}^{(1)}\gamma _{\lambda
}\gamma_5\nu^{(1)}~,~J_{\lambda}(\nu^{(1)})= \bar{\nu}^{(1)}\gamma
_{\lambda }(1-\gamma_5)\nu^{(1)}$$
 for Majorana and Dirac neutrinos respectively.
\item Process mediated by right handed currents  via Z'-boson exchange.\\
 The process is similar to that exhibited by Fig. \ref{fig:kkhz}, except that instead of Z we encounter Z', which is much
heavier.  Assuming that the couplings of the $Z'$ are similar
to those of $Z$, the above results apply except that now the
amplitudes are retarded by the multiplicative factor
$\kappa=m^2_{Z}/m^2_{Z'}$
\item Process mediated by Higgs exchange.\\
 In this case in Fig \ref{fig:kkhz} the Z is replaced by the Higgs particle.
 Proceeding as above we find that the amplitude at the nucleon level is:
  \beq
 {\cal M}_{\nu^{(1)}}(h)=
-2 \sqrt{2}G_F \frac{m_p m_{\nu^{(1)}}}{m_h^2}
\bar{\nu}^{(1)}~\nu^{(1)} \prec N|N \succ \sum_q f_q
 \eeq
  In the evaluation of the parameters
$f_q$ one encounters both theoretical and experimental errors.
\end{itemize}
\section{Other non SUSY Models}
\label{Zmodel}
 We should mention that there exist extensions of the standard model not motivated by symmetry. Such are:
\begin{itemize}
\item Models which introduce extra higgs particles and impose a discrete symmetry which leads to a "parity" 
a la R-parity or K-K parity \cite{MA06}.
\item Extensions of the standard model, which do not require a parity, but introduce high weak isospin
multiplets \cite{CFA06} with Y=0. So the WIMP-nucleus interaction via Z-exchange at tree level is absent and the dominant contribution to the WIMP-nucleus scattering occurs at the one loop level.
\item Another interesting extension of the standard model is in the direction of tecnicolor \cite{GKS06}. In this case the WIMP is 
the neutral LTP (lightest neutral technibaryon).
 This is scalar particle, which couples to the quarks via derivative coupling
through Z-exchange.
\end{itemize}
\section{Going from the Quark to the Nucleon Level} 
\label{sec:nuc}
 In going from the quark to the nucleon level one has to be a bit more 
careful in handling the quarks other than $u$ and $d$. This is especially true in the
case of the scalar interaction, since in this case the coupling 
of the WIMP to the quarks
is proportional to their mass~\cite{JDV06} .
Thus one has to consider in the nucleon not only
sea quarks ($u {\bar u}, d {\bar d}$ and $s {\bar s}$) but the heavier
quarks as well due to QCD effects ~\cite{Dree00} . 
This way one obtains the scalar Higgs-nucleon
coupling by using  effective parameters $f_q$ defined as follows:
\beq
\Big<N| m_q \bar{q}q|N \Big> = f_q m_N
\label{fofq}
\eeq
where $m_N$ is the nucleon mass. 
 The parameters $f_q,~q=u,d,s$ can be obtained by chiral
symmetry breaking 
terms in relation to phase shift and dispersion analysis (for a recent review see
\cite{JDV06}). We like to emphasize here that since the current masses of the u and d 
quarks are small, the heavier quarks tend to dominate even though the probability of
finding them in the nucleus is quite small. In fact the s quark contribution may become dominant,
e.g. allowed by the above analysis is the choice:
$$f_d=0.046,f_u=0.025,f_s=0.400,f_c=0.050,f_b=0.055,f_t=0.095$$

  The isoscalar and the isovector axial current in the case of K-K theories has already been discussed
  above. In the case of the neutralino these
couplings at the nucleon level, $ f^0_A$, $f^1_A$, are obtained from the corresponding ones given by the SUSY
 models at the quark level, $ f^0_A(q)$, $f^1_A(q)$, via renormalization
coefficients $g^0_A$, $g_A^1$, i.e.
$ f^0_A=g_A^0 f^0_A(q),f^1_A=g_A^1 f^1_A(q).$
The  renormalization coefficients are given terms of $\Delta q$ defined above \cite{JELLIS},
via the relations
$$g_A^0=\Delta u+\Delta d+\Delta s=0.77-0.49-0.15=0.13~,~g_A^1=\Delta u-\Delta d=1.26$$
We see that, barring very unusual circumstances at the quark level, the isoscalar contribution is
negligible. It is for this reason that one might prefer to work in the isospin basis.
\section{The allowed SUSY Parameter Space}
\label{sec:parameter}
 It is clear from the above discussion that the LSP-nucleon cross section depends, among other things,
on the parameters of supersymmetry.
 One starts with a set of parameters at the GUT scale and predicts the low energy
observables via the renormalization group equations (RGE). Conversely starting from the low energy phenomenology
one can constrain the input parameters at the GUT scale. 
 The parameter space is the most crucial. In SUSY models derived from minimal SUGRA
the allowed parameter space is characterized at the GUT scale  by five 
parameters:
\begin{itemize}
\item two universal mass parameters, one for the scalars, $m_0$, and one for the
fermions, $m_{1/2}$.
\item $tan\beta $, i.e the ratio of the Higgs expectation values, $\left < H_2 \right>/\left < H_1 \right>$.
\item The trilinear coupling  $ A_0 $ (or  $ m^{pole}_t $)  and 
\item The sign of $\mu $ in the Higgs self-coupling  $\mu H_1H_2$.
\end{itemize}
The experimental constraints \cite{JDV06} restrict the values of the above
parameters yielding the {\bf allowed SUSY parameter space}.
\section{Event rates}
\label{sec:rates}
The differential non directional  rate can be written as
\begin{equation}
dR_{undir} = \frac{\rho (0)}{m_{\chi}} \frac{m}{A m_N}
 d\sigma (u,\upsilon) | \mbox{\boldmath $\upsilon$}|
\label{2.18}
\end{equation}
where A is the nuclear mass number, 
$\rho (0) \approx 0.3 GeV/cm^3$ is the WIMP density in our vicinity,
 m is the detector mass, 
 $m_{\chi}$ is the WIMP mass and $d\sigma(u,\upsilon )$ is the differential cross section.
 
 The directional differential rate, i.e. that obtained, if nuclei recoiling in the direction $\hat{e}$ are 
observed, is given by \cite{JDVSPIN04,JDV06} :
\beq
dR_{dir} = \frac{\rho (0)}{m_{\chi}} \frac{m}{A m_N}
|\upsilon| \hat{\upsilon}.\hat{e} ~\Theta(\hat{\upsilon}.\hat{e})
 ~\frac{1}{2 \pi}~
d\sigma (u,\upsilon\
\nonumber \delta(\frac{\sqrt{u}}{\mu_r \upsilon
\sqrt{2}}-\hat{\upsilon}.\hat{e})
 \label{2.20}
\eeq
where $\Theta(x)$ is the Heaviside function.

The differential cross section is given by:
\beq
d\sigma (u,\upsilon)== \frac{du}{2 (\mu _r b\upsilon )^2}
 [(\bar{\Sigma} _{S}F(u)^2
                       +\bar{\Sigma} _{spin} F_{11}(u)]
\label{2.9}
\end{equation}
where $ u$ the energy transfer $Q$ in dimensionless units given by
\begin{equation}
 u=\frac{Q}{Q_0}~~,~~Q_{0}=[m_pAb]^{-2}=40A^{-4/3}~MeV
\label{defineu}
\end{equation}
 with  $b$ is the nuclear (harmonic oscillator) size parameter. $F(u)$ is the
nuclear form factor and $F_{11}(u)$ is the spin response function associated with
the isovector channel.

The scalar contribution is given by:
\begin{equation}
\bar{\Sigma} _S  =  (\frac{\mu_r}{\mu_r(p)})^2
                           \sigma^{S}_{p,\chi^0} A^2
 \left [\frac{1+\frac{f^1_S}{f^0_S}\frac{2Z-A}{A}}{1+\frac{f^1_S}{f^0_S}}\right]^2
\approx  \sigma^{S}_{N,\chi^0} (\frac{\mu_r}{\mu_r (p)})^2 A^2
\label{2.10}
\end{equation}
(since the heavy quarks dominate the isovector contribution is
negligible). $\sigma^S_{N,\chi^0}$ is the LSP-nucleon scalar cross section.

The spin contribution is given by:
\begin{equation}
\bar{\Sigma} _{spin}  =  (\frac{\mu_r}{\mu_r(p)})^2
                           \sigma^{spin}_{p,\chi^0}~\zeta_{spin},
\zeta_{spin}= \frac{1}{3(1+\frac{f^0_A}{f^1_A})^2}S(u)
\label{2.10a}
\end{equation}
\begin{equation}
S(u)\approx S(0)=[(\frac{f^0_A}{f^1_A} \Omega_0(0))^2
  +  2\frac{f^0_A}{ f^1_A} \Omega_0(0) \Omega_1(0)+  \Omega_1(0))^2  \, ]
\label{s(u)}
 \end{equation}
 The couplings $f^1_A$ ($f^0_A$) and the nuclear matrix elements $\Omega_1(0)$ ($\Omega_0(0)$) associated
 with the isovector (isoscalar) components are normalized so that, in the case
 of the proton at $u=0$, they yield $\zeta_{spin}=1$.

 With these definitions in the proton neutron representation we get:
 \beq
 \zeta_{spin}= \frac{1}{3}S^{'}(0)~,~S^{'}(0)=\left[(\frac{a_n}{a_p}\Omega_n(0))^2+2 \frac{a_n}{a_p}\Omega_n(0) \Omega_p(0)+\Omega^2_p(0)\right]
 \label{Spn}
 \eeq
 where $\Omega_p(0)$ and $\Omega_n(0)$ are the proton and neutron components of the static spin nuclear matrix elements. In extracting limits on the nucleon cross sections from the data we will find it convenient to
 write:
 \begin{equation}
                          \sigma^{spin}_{p,\chi^0}~\zeta_{spin} =\frac{\Omega^2_p(0)}{3}|\sqrt{\sigma_p}+\frac{\Omega_n}{\Omega_p} \sqrt{\sigma_n}
 e^{i \delta}|^2
\label{2.10ab}
\end{equation}
 In Eq. (\ref{2.10ab}) $\delta$ the relative
phase between the two amplitudes $a_p$ and $a_n$, which in most models is 0 or $\pi$, i.e. one expects them to be relatively real.
 The static spin matrix elements are obtained in the context of a given nuclear model. Some such matrix elements of interest to the planned experiments  can be found in \cite{JDV06}.
 
The spin ME are defined as follows:
\beq
\Omega_p(0)=\sqrt{\frac{J+1}{J}}\prec J~J| \sigma_z(p)|J~J\succ ~~,~~
\Omega_n(0)=\sqrt{\frac{J+1}{J}}\prec J~J| \sigma_z(n)|J~J\succ
\label{Omegapn}
\eeq
where $J$ is the total angular momentum of the nucleus and $\sigma_z=2 S_z$. The spin operator is defined by $S_z(p)=\sum_{i=1}^{Z} S_z(i)$, i.e. a sum over all protons in the nucleus,  and
$S_z(n)=\sum_{i=1}^{N}S_z(i)$, i.e. a sum over all neutrons. Furthermore $\Omega_0(0)=\Omega_p(0)+\Omega_n(0)~~,~~\Omega_1(0)=\Omega_p(0)-\Omega_n(0)$
\section{The WIMP velocity distribution}
To obtain the total rates one must fold with WIMP velocity distribution and integrate  the
above expressions  over the
energy transfer from $Q_{min}$ determined by the detector energy cutoff to $Q_{max}$
determined by the maximum LSP velocity (escape velocity, put in by hand in the
Maxwellian distribution), i.e. $\upsilon_{esc}=2.84~\upsilon_0$ with  $\upsilon_0$
the velocity of the sun around the center of the galaxy($229~Km/s$).

For a given velocity distribution f(\mbox{\boldmath $\upsilon$}$^{\prime}$),
 with respect to the center of the galaxy,
one can find the velocity distribution in the Lab
f(\mbox{\boldmath $\upsilon$},\mbox{\boldmath $\upsilon$}$_E$)
by writing 
\mbox{\boldmath $\upsilon$}$^{'}$=
          \mbox{\boldmath $\upsilon$}$ \, + \,$ \mbox{\boldmath $\upsilon$}$_E \, ,$
\mbox{\boldmath $\upsilon$}$_E$=\mbox{\boldmath $\upsilon$}$_0$+
 \mbox{\boldmath $\upsilon$}$_1$, with
\mbox{\boldmath $\upsilon$} $_1 \,$ the  Earth's velocity
 around the sun.
  
It is convenient to choose a coordinate system so  that
 $\hat{x}$  is radially out in the plane of the galaxy,
 $\hat{z}$ in the sun's direction of motion and 
 $\hat{y}=\hat{z}\times\hat{x}$.

Since the axis of the ecliptic  
lies very close to the $x,y$ plane ($\omega=186.3^0$) only  the angle
 $\gamma=29.8^0$
becomes relevant.
Thus the velocity of the earth around the
sun is given by 
\begin{equation}
\mbox{\boldmath $\upsilon$}_E  = \mbox{\boldmath $\upsilon$}_0 \hat{z} +
                                  \mbox{\boldmath $\upsilon$}_1  
(\, sin{\alpha} \, {\bf \hat x}
-cos {\alpha} \, cos{\gamma} \, {\bf \hat y}
+ cos {\alpha} \, sin{\gamma} \, {\bf \hat z} \,)
\label{3.6}  
\end{equation}
where $\alpha$ is  phase of the earth's orbital motion.

The WIMP velocity distribution f(\mbox{\boldmath $\upsilon$}$^{\prime}$) is not known. Many velocity distributions
have been used. The most common one is the M-B distribution with characteristic velocity $\upsilon_0$ 
with an upper bound $\upsilon_{esc}=2.84 \upsilon_0$. 
\beq
f(\mbox{\boldmath $\upsilon$}^{\prime})=\frac{1}{(\sqrt{\pi}\mbox{\boldmath $\upsilon_0$})^3}
e^{-(\mbox{\boldmath $\upsilon$}^{\prime}/\mbox{\boldmath $\upsilon_0$})^2}
\label{fv}
\eeq
Modifications of this velocity distribution
have also been considered such as: i) Axially symmetric M-B distribution  \cite {Druk,Verg00}. and ii) 
modifications of the characteristic parameters of the M-B distribution by considering a coupling
between dark matter and dark energy \cite{TETRVER06} 
($\upsilon_0 \rightarrow n \upsilon_0,\upsilon_{esc}\rightarrow n \upsilon_{esc}$). Other possibilities are adiabatic velocity distribution following the Eddington approach 
\cite{EDDIN}$^-$\cite{VEROW06} ,
caustic rings \cite{SIKIVI1}$^-$\cite{Gelmini} and  Sagittarius dark matter \cite{GREEN02} .
  
For a given energy transfer the velocity $\upsilon$ is constrained to be
\beq
\upsilon\geq \upsilon_{min}~,~\upsilon_{min}= \sqrt{\frac{ Q A m_p}{2}}\frac{1}{\mu_r}.
\eeq
\section{The Direct detection rate}
The event rate for the coherent WIMP-nucleus elastic scattering is given by \cite{Verg01,JDV03,JDVSPIN04,JDV06}:
\beq
R= \frac{\rho (0)}{m_{\chi^0}} \frac{m}{m_p}~
              \sqrt{\langle v^2 \rangle } \left [f_{coh}(A,\mu_r(A)) \sigma_{p,\chi^0}^{S}+f_{spin}(A,\mu_r(A))\sigma _{p,\chi^0}^{spin}~\zeta_{spin} \right]
\label{fullrate}
\eeq
with
\beq
f_{coh}(A, \mu_r(A))=\frac{100\mbox{GeV}}{m_{\chi^0}}\left[ \frac{\mu_r(A)}{\mu_r(p)} \right]^2 A~t_{coh}\left(1+h_{coh}cos\alpha \right)
\eeq
\beq
f_{spin}(A, \mu_r(A))=\left[ \frac{\mu_r(A)}{\mu_r(p)} \right]^2 \frac{t_{spin}(A)}{A}t_{spin}\left(1+h_{spin}cos\alpha \right)
\eeq
with $\sigma_{p,\chi^0}^{S}$ and $\sigma _{p,\chi^0}^{spin}$ the scalar and spin proton cross sections
$~\zeta_{spin}$ the nuclear spin ME. In the above expressions $h$ is the modulation amplitude.

 The number of events in time $t$ due to the scalar interaction, which leads to coherence, is:
\beq
 R\simeq 1.60~10^{-3}
\frac{t}{1 \mbox{y}} \frac{\rho(0)}{ {\mbox0.3GeVcm^{-3}}}
\frac{m}{\mbox{1Kg}}\frac{ \sqrt{\langle
v^2 \rangle }}{280 {\mbox kms^{-1}}}\frac{\sigma_{p,\chi^0}^{S}}{10^{-6} \mbox{ pb}} f_{coh}(A, \mu_r(A))
\label{scalareventrate}
\eeq
In the above expression
 $m$ is the target mass, $A$ is the number of nucleons
in the nucleus and $\langle v^2 \rangle$ is the average value of the square of the WIMP velocity.

In the case of the spin interaction we write:
\beq
 R\simeq 16
\frac{t}{1 \mbox{y}} \frac{\rho(0)}{ {\mbox0.3GeVcm^{-3}}}
\frac{m}{\mbox{1Kg}}\frac{ \sqrt{\langle
v^2 \rangle }}{280 {\mbox kms^{-1}}}\frac{\sigma_{p,\chi^0}^{S}}{10^{-2} \mbox{ pb}} f_{spin}(A, \mu_r(A))
\label{spineventrate}
\eeq
Note the different scale for the proton spin cross section.
The parameters $f_{coh}(A,\mu_r(A))$, $f_{spin}(A,\mu_r(A))$, which give the relative merit
 for the coherent and the spin contributions in the case of a nuclear
target compared to those of the proton,  have already been  tabulated \cite{JDV06}
 for energy cutoff $Q_{min}=0,~10$ keV. It is clear that for large A the coherent process is
expected to dominate unless for some reason the scalar proton cross section is very suppressed.

In the case of directional experiments the event rate is given
 by Eqs (\ref{scalareventrate}) and (\ref{spineventrate}) except that now:
 \beq
f_{coh}(A, \mu_r(A))=\frac{100\mbox{GeV}}{m_{\chi^0}}\left[ \frac{\mu_r(A)}{\mu_r(p)} \right]^2 A\frac{\kappa}{2 \pi}t_{coh}\left(1+h_m(coh)cos{(\alpha+\alpha_m \pi)} \right)
\eeq
\beq
f_{spin}(A, \mu_r(A))=\frac{100\mbox{GeV}}{m_{\chi^0}}\left[ \frac{\mu_r(A)}{\mu_r(p)} \right]^2 \frac{\kappa}{2 \pi}\frac{t_{spin}}{A}\left(1+h_m(spin)cos{(\alpha+\alpha_m \pi)} \right)
\eeq
In the above expressions $h_m$ is the modulation amplitude and $\alpha _m$ the shift in the phase of the modulation (in units of $\pi$) relative to the phase of the Earth. $\kappa/(2 \pi)$, $\kappa\leq 1$, is the suppression factor entering due to the restriction of
the phase space.  $\kappa$, $h_m$ and $\alpha_m$ depend on the direction of observation. It is precisely this
dependence as well as the large values of $h_m$, which can be exploited to reject background \cite{JDV06}, that makes the directional experiments quite attractive in spite of the suppression factor relative to the standard
experiments.
\section{Bounds on the scalar proton cross section}
\begin{figure}
\rotatebox{90}{\hspace{0.0cm} $\sigma_p\rightarrow 10^{-5}$pb}
\psfig{file=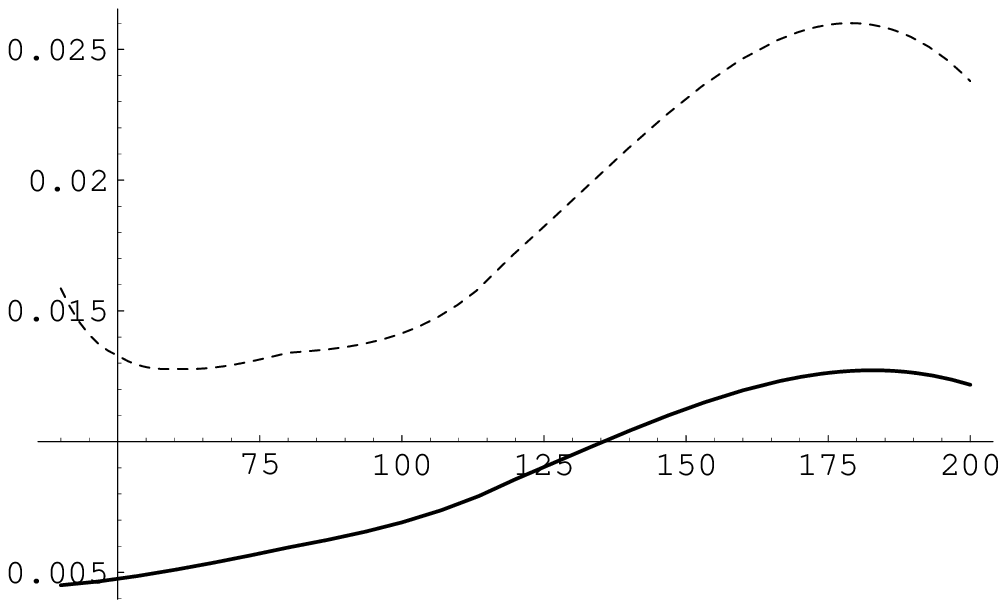,width=2.0in}
\rotatebox{90}{\hspace{0.0cm} $\sigma_p\rightarrow 10^{-5}$pb}
\psfig{file=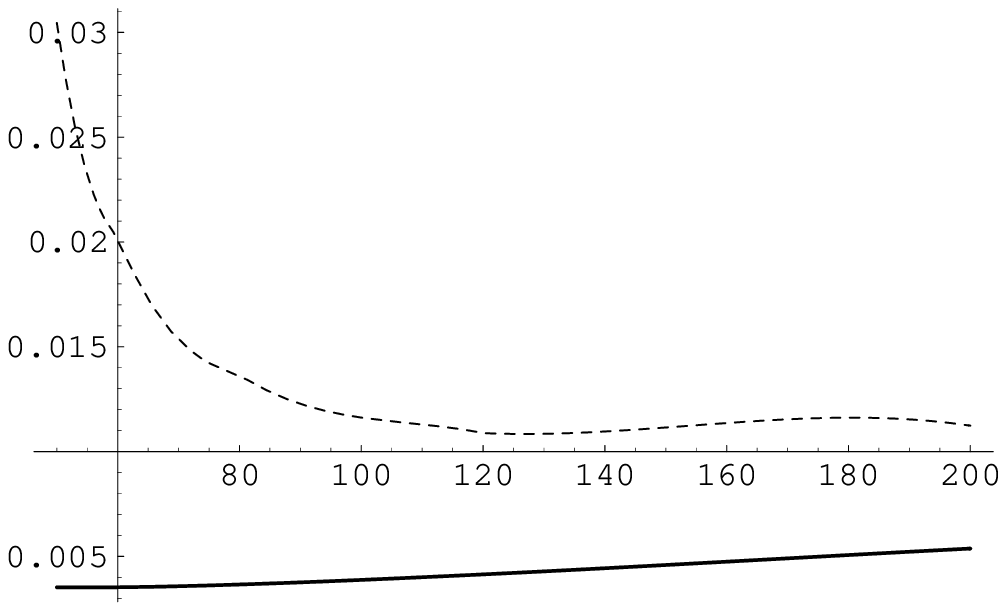,width=2.0in}
\hspace{-2.0cm} $m_{\chi}\rightarrow$ GeV
\caption{ The limits on the scalar proton cross section for A$=127$ on the left and A$=73$ on the right as functions of $m_{\chi}$. The continuous (dashed) curves correspond to $Q_{min}=0~(10)$ keV respectively. Note that the advantage of the larger nuclear mass number of the A$=127$ system is counterbalanced by the favorable form factor dependence of the A$=73$ system.}
 \label{b127.73}
\end{figure}

Using the above formalism one can obtain the quantities of interest $t$ and $h$ both for the standard as
well as the directional experiments. Due to lack of space we are not going to present the obtained results
here. The interested reader can find some of these results elsewhere \cite{JDVSPIN04,JDV06} . Here we are 
simply going to show how
one can employ such results to extract the nucleon cross section from the data.
Due to space considerations we are not going to discuss the limits extracted from the data on the spin cross
sections, since in this case one has to deal with two amplitudes (one for the proton and one for the neutron). We will only
extract some limits imposed on the
scalar nucleon cross section (the proton and neutron cross section are essentially the same).
In what follows we will employ for all targets \cite{BCFS02}$^-$\cite{PAVAN01} the limit of CDMS II for the Ge target \cite{CDMSII04} ,
 i.e.  $<2.3$ events for 
an exposure of $52.5$ Kg-d with a threshold of $10$ keV. This event rate is similar to that for other systems \cite{SGF05}. The thus obtained limits are exhibited in Fig. \ref{b127.73}. For larger WIMP masses one can
extrapolate these curves, assuming an increase as $\sqrt{m_{\chi}}$.
\section{Transitions to excited states}
The above formalism can easily be extended to cover transitions to excited states. Only the kinematics and 
the nuclear physics is different. In other words one now needs:
\begin{itemize}
\item The inelastic scalar form factor.\\
 The transition amplitude is non zero due to the momentum transfer involved. The relevant multipolarities
 are determined by the spin and parity of the final state.
 \item Spin induced transitions.\\
 In this case one can even have a Gamow-Teller like transition, if the final state is judiciously chosen.
 \end{itemize}
 
\begin{figure}
\begin{center}
\rotatebox{90}{\hspace{1.0cm} {\tiny BRR}$\rightarrow$}
\includegraphics[height=.17\textheight]{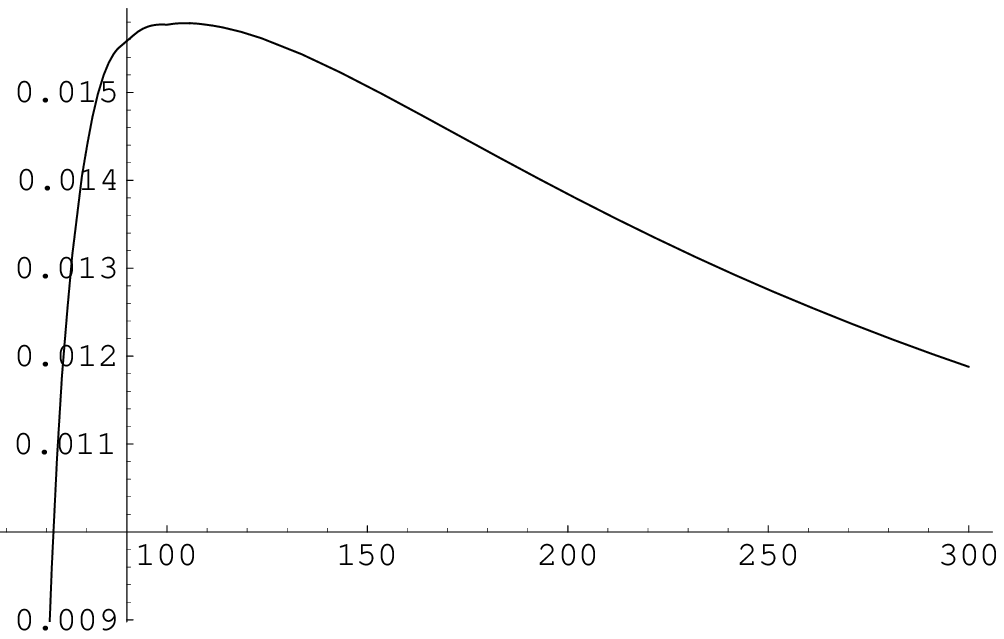}
 \rotatebox{90}{\hspace{1.0cm} {\tiny BRR}$\rightarrow$}
\includegraphics[height=.17\textheight]{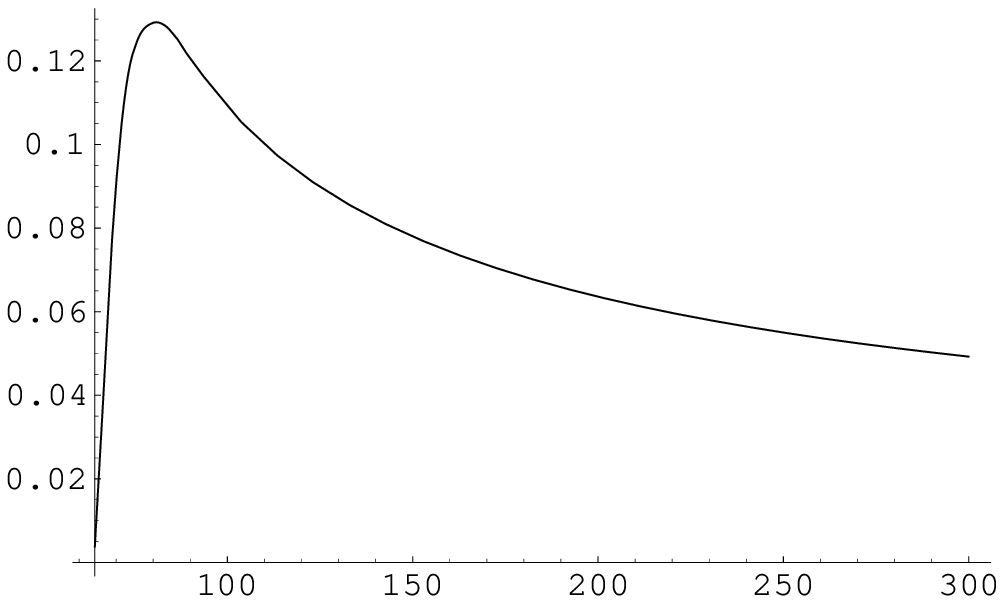}\\
 \hspace{0.0cm}$m_{LSP}\rightarrow$ ($GeV$)
 \caption{ The ratio of the rate to
the excited state divided by that of the ground state as a
function of the LSP mass (in GeV) for $^{127}I$.
 We found that
the static spin matrix element of the transition from the ground
to the excited state is a factor of 1.9 larger than that
involving the ground state and assumed that the spin response functions
$F_{11}(u)$ are the same. 
On the left we show the results for $Q_{min}=0$ and on
the right for $Q_{min}=10~KeV$. \label{ratio} }.
\end{center}
\end{figure}

 In the case of $^{127}I$ the static spin matrix element involving the first excited state around 
50 keV is twice as large compared to that of the
 ground state \cite{VQS04}. The spin response function was assumed to be the same with that of the ground
 state. The results obtained \cite{VQS04} are shown in Fig. \ref{ratio}.
These results are very encouraging, since, as we have mentioned, for heavier WIMPS like those involved
in K-K theories, the branching ratios are expected to be much larger. Thus one may consider such
transitions, since the detection of de-excitation $\gamma$ rays is much easier than the detection of
recoiling nuclei.
\section{Other non recoil experiments}
As we have already mentioned the nucleon recoil experiments are very hard. It is therefore necessary to consider
other possibilities. One such possibility is to detect the electrons produced during the WIMP-nucleus 
collisions \cite{VE05,MVE05} employing detectors with low energy threshold with a high Z target. Better yet
one may attempt to detect the very hard X-rays generated when the inner shell electron holes are 
filled \cite{EMV05}. The relative X-ray to nucleon recoil probabilities $[Z \sigma _K /\sigma _r]_i$, 
for $i=L (m_{\chi}\leq \mbox{100GeV}),~M(\mbox{100 GeV}\leq m_{\chi}\leq  \mbox{200 GeV})$ and $H (m_{\chi}\simeq \mbox{200 GeV})$ are shown in table \ref{table:X-rays}. For even heavier WIMPs, like
those expected in K-K theories, the relative probability is expected to be even larger.

\begin{table}
\begin{center}
\caption{K X-ray cross sections relative to the nuclear recoil,
rates and energies in WIMPs nuclear interactions with $^{131}$Xe.
$[Z \sigma _K /\sigma _r]_L, [Z \sigma _K /\sigma _r]_M$ and $[Z
\sigma _K /\sigma _r]_H$ are the ratios for light (30 GeV), medium
(100 GeV) and heavy (300 GeV) WIMPs.} \label{t:2} \vspace{0.5cm}
\label{table:X-rays}
\begin{tabular}{|ccccc|}
\hline K X-ray & $E_K(K_{ij})$ keV    & $[\frac{Z
\sigma _K(K_{ij})}{\sigma _r}]_{L}$ &  $[\frac{Z \sigma
_K(K_{ij})}{\sigma _r} ]_{M} $ & $[\frac{Z \sigma
_K(K_{ij})}{\sigma _r}]_{H} $ \\ \hline
 K$_{\alpha 2}$  & 29.5  & 0.0086 & 0.0560 & 0.0645 \\
 K$_{\alpha 1}$ & 29.8  & 0.0160 & 0.1036 & 0.1196 \\
 K$_{\beta 1}$  & 33.6   & 0.0047 & 0.0303 & 0.0350 \\
 K$_{\beta 2}$  & 34.4   & 0.0010 & 0.0067 & 0.0077 \\

\hline
\end{tabular}
\end{center}
\end{table}

The K$_{\alpha}$ and K$_{\beta}$ lines can be
separated experimentally by using good energy-resolution
detectors, but the sum of all K lines can be measured in modest
energy-resolution experiments.
\section{Conclusions}
We examined the various signatures expected in the direct detection of WIMPs via their
interaction with nuclei. We specially considered WIMPs predicted in supersymmetric models (LSP or neutralino) as well as  theories with extra dimensions. We presented the formalism for the modulation 
amplitude for non directional as well as directional experiments. We discussed the role played by
nuclear physics on the extraction of the nucleon cross sections from the data. We also considered
non recoil experiments, such as measuring the $\gamma$ rays following the de-excitation of the nucleus
and/or the hard X-rays after the de-excitation of the inner shell electron holes produced during the WIMP 
nucleus interaction. These are favored by very heavy MIMPs in the TeV region and velocity distributions
expected in models allowing interaction of dark matter and dark energy.

{\bf Acknowledgments}:
This work was supported in part by the European Union contract MRTN-CT-2004-503369.  Special thanks to Professor Raduta for support and hospitality during  the Predeal Summer School.

\end{document}